\documentstyle[12pt,epsf]{article}

\newcommand{\be}{\begin{equation}}
\newcommand{\ee}{\end{equation}}
\newcommand{\bea}{\begin{eqnarray}}
\newcommand{\eea}{\end{eqnarray}}
\newcommand{\ra}{\rightarrow}
\newcommand{\lan}{\langle}
\newcommand{\ran}{\rangle}
\newcommand{\YY}{Y\hspace{-2.75mm}Y}
\newcommand{\LL}{I\hspace{-1.65mm}L}
\newcommand{\PP}{I\hspace{-1.65mm}P}
\newcommand{\hh}{h\hspace{-1.85mm}h}
\begin{document}

\font\fortssbx=cmssbx10 scaled \magstep2
\hbox to \hsize{
\includegraphics{uwlogo.ps}
\hskip.5in \raise.1in\hbox{\fortssbx University of Wisconsin - Madison}
\hfill$\vcenter{\hbox{\bf MADPH-94-852}
            \hbox{November 1994}}$ }
\vskip 2cm
\begin{center}
\Large
{\bf Observations On the Potential Confinement of a Light Fermion} \\
\vskip 0.5cm
\large
M. G. Olsson and  Sini\v{s}a Veseli \\
\vskip 0.1cm
{\small \em Department of Physics, University of Wisconsin, Madison,
	\rm WI 53706} \\
\vspace*{+0.2cm}
 Ken Williams \\
{\small \em Continuous Electron Beam Accelerator Facility \\
	Newport News, VA 29606, USA \\
	and \\
\vspace*{-0.2cm}
Physics Department, Hampton University, Hampton, VA 29668}
\end{center}
\thispagestyle{empty}
\vskip 0.7cm
\begin{abstract}
We consider possible dynamical models
for a light fermion confined by a potential field.
With the Dirac equation only  Lorentz scalar
confinement
yields normalizable wavefunctions, while with the
``no pair'' variant of the Dirac equation
only Lorentz vector confinement has normal Regge behaviour.
 A systematic
investigation of Regge properties
and phenomenological properties is carried out, including
calculations of the Isgur-Wise function. We point out that
the Isgur-Wise function provides a sensitive
test of  confinement models. In particular, the slope
of the IW function at zero recoil point
is found to be $\xi'(1)\simeq -0.90$ for the Dirac equation
with scalar confinement, and $\xi'(1)\simeq -1.20$ for the
no pair equation with vector confinement. Using heavy-light data
alone we argue against scalar confinement.
\end{abstract}

\newpage
\section{Introduction}
The description of the dynamical confinement of a light fermion
by a central force requires an examination
of various wave equations. For example, the direct application
of the Dirac equation with Lorentz vector confinement
leads to pair production on hadronic time scales which is not observed.
In this case the ``no pair'' (NP) or ``quenched'' variant of the Dirac
equation is used to obtain a physical and ``well posed'' result.

In this paper we survey the solutions of the Dirac and NP equation for both
 Lorentz vector and scalar confining potentials. We use the
results in each case to compare to heavy-light meson data. Dirac scalar
confinement and NP vector confinement are
capable of accounting for the measured meson masses
as well as the Isgur-Wise function. The values of the parameters
resulting from the fits however provide interesting clues to the
correct
form of the confinement. In particular, we will argue
against scalar confinement on the basis of heavy-light data alone.

We begin in Section \ref{sec:weqs} with a brief discussion
of one particle fermionic wave equations and their properties. Classical
arguments reveal the Regge behaviour and categorize the effects
of the negative energy states on the positive
energy solutions. In Section \ref{sec:nmeth} we discuss
our numerical method and reproduce the pure Coulomb NP
solution of Hardekopf and Sucher \cite{bib:hardekopf}. We then
successivly consider scalar and vector confined mesons
with Dirac and NP equations in Section \ref{sec:dnp} and we
summarize our findings in Section \ref{sec:conc}.

\section{Wave equations}
\label{sec:weqs}

The time independent Dirac equation for a  quark of mass
$m$ moving in a Lorentz scalar potential $S(r)$ and the
time component of a Lorentz vector
potential $V(r)$ is
\be
[H_{0}+ P   - E]\psi({\bf r})=0\ ,\label{eq:deq}
\ee
where
the free particle Dirac Hamiltonian is
\be
H_{0} = \mbox{\boldmath $\alpha$}\cdot {\bf p} +\beta m\ ,
\ee
and
\be
P(r)  = V(r)+\beta S(r)\ .
\ee
 E is the energy of the ``light degrees of freedom'',
and \mbox{\boldmath $\alpha$} and $\beta$ are the usual Dirac matrices
\cite{bib:gross}.

When one attempts to use the Dirac equation to solve the Helium
atom
 one finds \cite{bib:hardekopf,bib:brown}
that no normalizable solutions are possible. This ``continuum
dissociation'' is a form of the Klein paradox when the negative
energy states mix with the positive energy states.
Since Helium clearly exists,
one  must rephrase the Dirac equation to suppress this mixing. The result is
the NP equation \cite{bib:hardekopf}.

An analogous phenomenon
occurs for a single particle moving in
an increasing (confining) Lorentz vector potential.
For a very long time \cite{bib:plesset} it
has been realized that
there are no normalizable solutions to the Dirac equation in this case.
This again is an example of the
Klein paradox \cite{bib:bjorken}. For a potential which rises
sufficiently fast and a sufficiently small quark mass,
the states corresponding to the free
negative and positive energies  mix resulting in
the loss of normalizability. In both these cases
a simple alternation to the Dirac equation
avois this mixing. The free Dirac Casimir projection
operators \cite{bib:gross} are
\be
\Lambda_{\pm}=\frac{E_{0}\pm H_{0}}{2E_{0}}\ ,
\ee
where
\be
E_{0}=\sqrt{{\bf p}^{2}+m^{2}}\ .
\ee
These projection operators have the well known
properties,
\bea
&\Lambda_{\pm}^{2} = \Lambda_{\pm}\ ,&  \nonumber  \\
&\Lambda_{+}\Lambda_{-}= \Lambda_{-}\Lambda_{+} = 0\ ,& \nonumber \\
&\Lambda_{+} + \Lambda_{-} = 1 \ ,& \label{eq:lis}\\
&\Lambda_{\pm}H_{0}=H_{0}\Lambda_{\pm}\ .\nonumber
\eea

If $\Lambda_{+}$ acts on the Dirac equation (\ref{eq:deq})
from the left we obtain by use of (\ref{eq:lis})
\be
\Lambda_{+}(H_{0}+P  - E)\psi =[H_{0}\Lambda_{+}+
\Lambda_{+} P (\Lambda_{+}
+\Lambda_{-}) - E\Lambda_{+}]\psi = 0\ .
\ee
The $\Lambda_{-}\psi$ term represents  pair production
in the
interaction \cite{bib:hardekopf}.
The NP  approximation is obtained
by dropping the $\Lambda_{-}\psi$ term. Defining
a new wave
function
\be
\phi = \Lambda_{+}\psi\ ,\label{eq:phidef}
\ee
the resulting NP equation is
\be
(H_{0}+P_{+}-E)\phi = 0\ ,\label{eq:npeq}
\ee
where
\be
P_{+}= \Lambda_{+}P \Lambda_{+}\ .
\ee

\subsection{Classical turning points}

The s-wave classical turning
points
provide valuable insight into the
structure of these wave equations. By inspection, some vital properties
of the solutions will be evident. The s-wave turning
points are defined by ${\bf p}=0$ in  (\ref{eq:deq}) and
(\ref{eq:npeq}). The turning point
condition for the Dirac equation (\ref{eq:deq}) is
\be
\beta(m+S)+V = E\ .
\ee
There are two turning points; one bounding the positive
energy states $(\beta=1)$ and the other bounding
the negative energy states when $\beta = -1$. They are
\bea
E_{+} &=&  m + S(r) + V(r)\ ,\label{eq:tp1}\\
E_{-} &=& -m - S(r) + V(r)\ .\label{eq:tp2}
\eea
We consider in turn scalar confinement $[V=0,\ S=ar]$
and vector confinement $[V=ar, \ S=0]$. The s-wave
turning points given in  (\ref{eq:tp1}) and (\ref{eq:tp2})
are plotted in Figs. \ref{fig:tps} and \ref{fig:tpv}
for scalar and vector confinement respectively.
The two cases are dramatically
different. The scalar confinement
turning points insure separation of the positive and
negative state regions resulting in a well defined mathematical
problem. The vector confinement turning points
plotted in Fig. \ref{fig:tpv} allow
the free negative energy states to rise into the
positive
energy region causing mixing for a given positive energy.
This mixing is
known as the Klein paradox \cite{bib:bjorken}.
In principle this means there are no normalizable
bound states although for heavy quarks
the mixing effect is negligible.

At the s-wave turning point
we have
\be
\Lambda_{+}({\bf p}=0) = \left(\begin{array}{cc}
		1 & 0 \\
		0 & 0 \end{array}\right)\ ,
\ee
and the NP turning points from (\ref{eq:npeq}) are given by
\bea
E_{+} &=&  m + S(r) + V(r)\ ,\label{eq:tp3}\\
E_{-} &=& -m \ .\label{eq:tp4}
\eea
These are plotted in Fig. \ref{fig:tpnp} and are the same for
scalar and vector confinement. From this point of view
vector confinement
has become
a well defined concept since normalizable bound states
are now expected to exist.

\subsection{Spherical solutions}

Because of  spherical symmetry, we look for the
solution in the form
\cite{bib:gross}
\bea
\Psi_{jm}^{k}({\bf r})&=&
			\left(\begin{array}{c}
			f_{j}^{k}(r){\cal Y}_{jm}^{k}({\bf \hat{r}}) \\
			ig_{j}^{k}(r){\cal Y}_{jm}^{-k}({\bf \hat{r}})
			\end{array}\right)\ \nonumber \\
&\equiv& \YY\left(\begin{array}{c}
  		f_{j}^{k} \\
		g_{j}^{k} \end{array}\right) \ ,
\eea
where
\be
\YY = \left(\begin{array}{cc}
	{\cal Y}_{jm}^{k} & 0 \\
	0 		  & i{\cal Y}_{jm}^{-k}\end{array}\right)\ ,
\ee
and ${\cal Y}_{jm}^{k}$ are the spherical spinors.
Using the identity
\be
\mbox{\boldmath $\sigma$} \cdot {\bf p} {\cal Y}_{jm}^{\pm k}
= i{\cal Y}_{jm}^{\mp k}D_{\pm}\ ,
\ee
where
\be
D_{\pm} = \pm \frac{k}{r} + (\frac{d}{dr} + \frac{1}{r})\ ,\label{eq:dmindpl1}
\ee
and
\be
D_{-}=-D_{+}^{\dagger}\ ,\label{eq:dmindpl}
\ee
which can be verified by integration by parts.
We commute the $\bf Y$ matrix to the left
to obtain the standard coupled radial equations,
\be
(\hh_{0}+ \PP  -E{\bf{1}})\left(\begin{array}{c}
  		f_{j}^{k} \\
		g_{j}^{k} \end{array}\right) = 0\ ,\label{eq:rdeq}
\ee
where
\bea
\hh_{0}&=&\left(\begin{array}{cc}
		m & -D_{-}\\
		D_{+} & -m \end{array}\right)\ ,\\
 \PP  &=&\left(\begin{array}{cc}
		V+S & 0\\
		0 & V-S \end{array}\right)\ .
\eea
The quantum number $k$ labels the meson
and is defined by
\be
k=\pm(j+\frac{1}{2})\ ,\label{eq:defk}
\ee
where the $+$ sign means $l=j+\frac{1}{2}$ or $k=l$,
 the $-$ sign means $l=j-\frac{1}{2}$ or $k=-(l+1)$, and hence
$l(l+1) = k(k+1)$.

An analogous pair
of  coupled first order
radial equations, can be obtained for the NP
equation (\ref{eq:npeq}). We define
\be
\lambda_{\pm}=\frac{E_{0}^{\pm k}\pm m}{2 E_{0}^{\pm k}}	\ ,
\label{eq:lpllmin}
\ee
where
$E_{0}^{\pm k}$ is to be evaluated
using orbital
angular momentum $l$ according to (\ref{eq:defk}), i.e.
\be
E_{0}^{k}=\sqrt{p_{r}^{2}+\frac{k(k+1)}{r^{2}}+m^{2}}\ .\label{eq:e0}
\ee
The identity
\be
\Lambda_{+}\YY = \YY\LL\ ,
\ee
with
\be
\LL = \left(\begin{array}{cc}
		\lambda_{+} & -\frac{1}{2E_{0}^{k}}D_{-}\\
		D_{+}\frac{1}{2E_{0}^{k}} & \lambda_{-} \end{array}\right)\ ,
\label{eq:ll}
\ee
then gives the radial NP equation
\be
(\hh_{0}+\LL \PP  \LL-E{\bf{1}})\left(\begin{array}{c}
  		f_{j}^{k} \\
		g_{j}^{k} \end{array}\right) = 0\ .\label{eq:rnpeq}
\ee
It should be noted from  (\ref{eq:dmindpl}) and (\ref{eq:ll}) that
$\LL^{\dagger}=\LL$.

The radial equations for the Dirac case (\ref{eq:rdeq})
or the NP case (\ref{eq:rnpeq}) are solved in a similar manner
as discussed in Section \ref{sec:nmeth}. The NP equation
can of course
be thought of as a Dirac equation
with a coordinate and momentum dependent potential.

\subsection{Regge behaviour}
\label{sec:regge}

In the ultra-relativistic limit one aspect of the Dirac solutions can be
inferred
immediately without any detailed calculation. For large orbital
angular momenta the Regge trajectories become linear for
linear confinement
and have
a slope that is characteristic of the Lorentz nature of the confinement.
The result that the single light quark Regge slopes are exactly
double the corresponding two light quark Regge
slopes should be noted \cite{bib:goebel}, where in each case the energy
in the Regge slope is identified as the excitation energy (i.e., meson
energy minus heavy quark mass).

We consider high rotation $|k|\gg 1$
and nearly circular orbits. The radial Dirac equation (\ref{eq:rdeq})
then implies
\be
-\frac{g^{k}_{j}}{f^{k}_{j}}=
\frac{V+S-E}{\frac{k}{r}}=\frac{\frac{k}{r}}{V-S-E}\ ,
\ee
which becomes
\be
(E-V)^{2}-S^{2}=\frac{k^{2}}{r^{2}}\ .
\ee
The state of the lowest
energy for a fixed $k$
satisfies $\frac{\partial E}{\partial r}|_{k} = 0$, or
\be
(E-V)^{2}V'+SS'=\frac{k^{2}}{r^{3}}\ .
\ee
The two cases we are considering are
\begin{itemize}
\item Scalar confinement: $S= ar,\ V=0$, for which
\be
\alpha' = \frac{|k|}{E^{2}}=\frac{1}{2a}\ .\label{eq:scsl}
\ee
\item Vector confinement: $S=0,\ V=ar$, giving a  Regge slope of
\be
\alpha' = \frac{|k|}{E^{2}}=\frac{1}{4a}\ .\label{eq:vecsl}
\ee
\end{itemize}

The Regge behaviour of the NP equation proceeds similarly.
By (\ref{eq:dmindpl1}), (\ref{eq:lpllmin})  and (\ref{eq:e0})
the high $|k|\gg 1$ limit
implies
that
\be
\LL(|k|\gg 1) = \frac{1}{2}\left(\begin{array}{cc}
	1 & \hat{k} \\
	\hat{k} & 1 \end{array}\right)\ ,
\ee
where $\hat{k}=\frac{k}{|k|}$, and the NP equation potential term
is
\bea
\LL \PP  \LL &\simeq& \frac{1}{4}
\left(\begin{array}{cc}
	1 & \hat{k} \\
	\hat{k} & 1 \end{array}\right)
\left(\begin{array}{cc}
	V+S & 0 \\
	0 & V-S \end{array}\right)
\left(\begin{array}{cc}
	1 & \hat{k} \\
	\hat{k} & 1 \end{array}\right)\nonumber \\
&=&\frac{1}{2}V(r)\left(\begin{array}{cc}
	1 & \hat{k} \\
	\hat{k} & 1 \end{array}\right)\ .\label{eq:lpl}
\eea

We first note that the scalar potential cancels in this limit
and hence
the universal Regge behaviour is lost with
NP scalar confinement. The radial equation (\ref{eq:rnpeq})
then yields
\be
-\frac{g^{k}_{j}}{f^{k}_{j}}=
\frac{-E+\frac{1}{2}V}{\hat{k}(\frac{|k|}{r}+\frac{1}{2}V)}
=\frac{\hat{k}(\frac{|k|}{r}+\frac{1}{2}V)}{-E+\frac{1}{2}V}\ .
\ee
The solutions are $E=\frac{-|k|}{r}$ and a positive solution
\be
E=V(r)+\frac{|k|}{r}\ .
\ee
For the positive solution the minimum $E$ for a fixed $k$
condition then gives
\be
\alpha' = \frac{|k|}{E^{2}}=\frac{1}{4a}\ .\label{eq:npsl}
\ee
This is identical to the Dirac vector confinement slope
(\ref{eq:vecsl}).

We note that although
the origin of the NP equation lies in nearly
non-relativistic atomic physics it retains
vector confinement in the ultra-relativistic limit. The status of
scalar confinement
is drastically different. The situation here is reminiscent of the
scalar confinement
catastrophe which occurs in the momentum space formulation \cite{bib:gara}
where confinement is lost.

\section{Description of the numerical method}
\label{sec:nmeth}

Instead of solving (\ref{eq:rdeq})
by integrating differential equations, we choose a variational (Galerkin)
method \cite{bib:jacobs}. We expand $f_{j}^{k}$ and $g_{j}^{k}$ in terms of
a complete set of basis states $\phi(r)$, and
truncate the expansion to the lowest $N$ basis functions, i. e.
\be
f_{j}^{k}(r)\simeq\sum_{n=1}^{N}c_{n}^{(f)}\phi_{nl(k)}\ ,
\ee
and
\be
g_{j}^{k}(r)\simeq\sum_{n=1}^{N}c_{n}^{(g)}\phi_{nl(k)}\ .
\ee
Substituting these expressions into  (\ref{eq:rdeq}) or (\ref{eq:rnpeq}),
and then multiplying from the left by
\be
\int r^2 dr \phi_{ml(k)}\ ,
\ee
we end up with $2N\times 2N$ matrix equation in the form
\be
E\left(\begin{array}{c} c^{f} \\
			c^{g} \end{array} \right)
=
\left(\begin{array}{cc} H_{11} & H_{12} \\
			H_{21} & H_{22} \end{array} \right)
\left(\begin{array}{c} c^{f} \\
			c^{g} \end{array} \right)
=H\left(\begin{array}{c} c^{f} \\
			c^{g} \end{array} \right)
\ .
\ee
Here, $H_{ij}$ are symmetric $N\times N$ matrices, and
$H_{12}=H_{21}^{\dagger}$. Diagonalizing the
Hamiltonian matrix yields energies and eigenvectors in terms
of the basis states. The lowest $N$ energies and eigenvectors correspond
to  negative energy states, while the $N+1$ to $2N$ states
describe  positive energy states. Basis states, and therefore
energies and eigenvectors of $H$,
depend on the variational size parameter $\beta$. The
dependence on $\beta$
 of the
lowest positive energy  states (as well as the
 highest negative energy states) should vanish
with increasing
number $N$ of basis states used. The pseudo-Coulombic
basis states and all matrix
elements used are described in Appendix \ref{app:bstates}.

The Galerkin finite basis method will approach
the true eigenvalues from above if the Hamiltonian matrix
is positive definite. Although
the Hamiltonians which we are considering
here are not positive definite
they still exhibit plateaus
in $\beta$, which become  wider
as the number of basis states increases. In those cases where
analytic or alternative numerical solutions are available our Galerkin
plateaus correspond to the correct eigenvalues.

In order to see how the variational method works with the Dirac equation,
we first solve pure  Coulomb potential,
\bea
S(r)&=&0\ ,\\
V(r)&=&-\frac{\kappa}{r}\ ,
\eea
for which the analytical solution is known \cite{bib:gross}.
In Fig. \ref{fig:cbplt}
we show dependence of the three lowest positive energy states
(for $m=1\ GeV$, $\kappa=0.5$, $k=-1$ and $j=\frac{1}{2}$)
on the variational parameter $\beta$. The full lines represent
the exact analytic solution of the Dirac equation. As we increase
the number of basis states, the plateau region of $\beta$
where the eigenvalues of $H$
are the same as the exact energies enlarges. The variational scheme works
well in this case.

 In Fig. \ref{fig:kplt} we
illustrate the scaled energy of the ground state,
\be
\varepsilon =  \frac{E-m}{\kappa^{2}m}\ ,
\ee
as a function of the Coulomb constant $\kappa$. The solid line shows
the exact analytic Dirac result,
\be
E=m\sqrt{1-\kappa^{2}}\ ,
\ee
and also our numerical solution which are the same to high accuracy, even
for values of $\kappa$ close to one.
The dashed
line is the Coulomb NP equation ground state energy. As Hardekopf
and Sucher \cite{bib:hardekopf} found the NP
and Dirac solution are nearly the same for small $\kappa$. Our NP
Coulomb solution is consistent
with that obtained by  Hardekopf
and Sucher \cite{bib:hardekopf}.

\section{Results}
\label{sec:dnp}

\subsection{Dirac equation with scalar confinement}
Properties of the spectrum of the
Dirac equation with scalar confinement and a short
range Coulomb interaction have already been investigated
in \cite{bib:mur}. These authors also report results of their
numerical calculations,
which we have used as  another check of our method. With
$N=25$ basis states we were
able to reproduce  all of their numerical results
for the eigenvalues to their given accuracy of four decimal places.

As in \cite{bib:mur}, we take
\bea
S(r)&=&ar\ ,\nonumber \\
V(r)&=&-\frac{\kappa}{r}\ ,\label{eq:sconf}
\eea
but here we go a little bit further  in investigating the
use of the Dirac equation with scalar confinement in the
description of heavy-light mesons. First, we make sure
that nothing in our final result
 depends on the value of variational prameter $\beta$.
In Fig. \ref{fig:scbplt} we show the three lowest positive energy
states and the three highest negative energy states for
$m_{q}=0.3 GeV,\ a=0.2\ GeV^{2},\ \kappa=0.5,\ k=-1$ and
$j=\frac{1}{2}$. Clearly, with $N=15$ we have a large
region where the eigenvalues do not depend on $\beta$.

Next, we perform a
systematic fit to the observed spin averaged heavy-light
meson states. We fix the light quark mass
to be $m_{u,d}=0.3\ GeV$, and vary all the
other parameters of the model to best account
for the experimental data. In Table \ref{tab:sconf} we show the results
of this fit, with parameters
\bea
m_{u,d}&=&0.300 \ GeV \ {\rm (fixed)}\ ,\nonumber \\
m_{s}&=&0.463\ GeV\ ,\nonumber \\
m_{c}&=&1.301\ GeV\ ,\nonumber \\
m_{b}&=&4.639\ GeV\ ,\label{eq:scfit}\\
a&=&0.308 \ GeV^{2}\ ,\nonumber \\
\kappa&=&0.579\ . \nonumber
\eea
As seen from the Table \ref{tab:sconf},
the agreement with experimental data is excellent, and
values of parameters are all reasonable, except for
the value of the  tension $a$.

{}From the universal Regge slope
$\alpha' \simeq 0.8\ GeV^{-2}$, one expects $a$ to be
\be
a=\frac{1}{2\pi \alpha'}\simeq 0.2 GeV^{2}\ ,\label{eq:nambu}
\ee
and this is consistent with the value found from analyses of heavy onia
spectroscopies \cite{bib:jacobs}.
 However, as we saw in (\ref{eq:scsl}),  the
Regge slope
for the Dirac equation with scalar confinement is
$\alpha'=\frac{1}{2a}$, so that   tension
necessary to account for the spin averaged meson masses
 must be about a factor of $\frac{\pi}{2}$ larger than the one
expected from light-light  spectroscopies and the Nambu
slope (\ref{eq:nambu}). Indeed, if we
divide value of $a$ from  (\ref{eq:scfit}) by $\frac{\pi}{2}$, we
get 0.2, as we expected.

As we have already mentioned, Regge slope for the
Dirac equation with scalar confinement
is expected to be $\alpha'=\frac{1}{2a}$. In order to
verify this numerically, we fix
$a$ to be $0.2\ GeV^{2}$, choose   $m_{q}=0$, and then
plot dependence of $j$ with respect to  $\frac{E^{2}}{2a}$. As
seen on the Fig. \ref{fig:scregge}, slope of the
Regge trajectories is one, as expected.

Once the wave functions are known for a heavy light meson
the Isgur-Wise function, describing
the semi-leptonic $\bar{B}\ra D^{(*)}$
decay distribution, can be evaluated \cite{bib:sadz,bib:sik2}.
Using
\be
\xi(\omega)
=
\frac{2}{\omega + 1}
\lan\
j_{0}(2E_{q}\sqrt{\frac{\omega-1}{\omega+1}}r)
\ran\ , \label{eq:iwf}
\ee
where
\be
\lan A \ran =
\int_{0}^{\infty}dr\  r^2 R(r)A(r)R(r)\ ,
\ee
we find IW function that this model predicts. As shown on
the Fig. \ref{fig:sciwf}, the agreement with
ARGUS \cite{bib:arg} and CLEO \cite{bib:cleo} data is reasonable.
To calculate the slope, we use expression \cite{bib:sik2}
\be
\xi'(1) = -(\frac{1}{2}+\frac{1}{3}E_{q}^{2}
\lan r^{2}\ran) \ ,\label{eq:first}\\
\ee
For the range of light quark masses from 0 to 350 $MeV$,
we obtain
\be
\xi'(1)=-0.90\pm0.02\ .\label{eq:scslope}
\ee

\subsection{Dirac equation with vector confinement}

If a fermion is confined by a Lorentz vector interaction, i. e.
\bea
S(r)&=&0\ ,\nonumber \\
V(r)&=&-\frac{\kappa}{r}+ar\ ,\label{eq:vconf}
\eea
the Dirac equation has no normalizable solutions. As we mentioned earlier
this has been known to be the case for over sixty years \cite{bib:plesset}.
The origin of the problem
is in the mixing between
positive and negative energy states as shown
in Fig. \ref{fig:tpv}. Since the basis wavefunctions are
all normalizable in our variational method it is of interest
to see how
this problem is manefest. In Fig. \ref{fig:vcbplt}
we show the $\beta$-plot
with vector
confinement for the three lowest positive and the three highest
negative energy states. We see how the three states mix
in the region of $\beta$ where there should be a plateau.
It is interesting to note however that if one extends
the apparent plateaus through the level repulsions
a consistent result
is obtained which is not too different from
the NP result below. The largest difference is
in the ground state where Dirac and NP are about $50\ MeV$ apart.

\subsection {NP equation with scalar confinement}
We observed in (\ref{eq:lpl}) that for large orbital
excitation
the NP scalar interaction cancels
from the NP equation. The consequent loss of linear Regge
trajectories  eliminates any conventional discussion
of meson states in terms of scalar confinement in the NP framework.

\subsection{NP equation with vector confinement}

There is  reason to hope that the NP equation with
vector confinement (\ref{eq:vconf}) will eliminate this mixing and
hence
reestablish  a one particle wave equation. Comparing the s-wave
turning point structure of the NP equation in Fig. \ref{fig:tpnp}
with the Dirac equation shown in Fig. \ref{fig:tpv} we
observe
that the NP negative energy states should not seriously mix with
the positive
energy states.

We proceed to the numerical solution of the NP
equation (\ref{eq:rnpeq}) by the Galerkin method.
 In Fig. \ref{fig:npbplt}
we show the dependence on $\beta$ of the three lowest
positive and three highest negative energy solutions with
$m_{q}=0.3\ GeV,\ a=0.2\ GeV^{2},
\ \kappa=0.5,\ k=-1$ and $j=\frac{1}{2}$. Again we observe
a robust plateau structure which
widens as the number of basis functions increases.

The Regge slope of the NP equation with vector confinement
was shown in (\ref{eq:npsl}) to be $\alpha'=\frac{1}{4a}$. Our
numerical solution agrees as shown
in Fig. \ref{fig:npregge}. In this figure we illustrate
the leading trajectories for
the two light degrees of freedom
states
corresponding to $k=\pm(j+\frac{1}{2})$ for  $m_{u,d}=0.3\ GeV,
\ a=0.2\ GeV^{2}$ and $\kappa = 0.5$. We also
show several
daughter trajectories
corresponding to radial excitations.

In order to recover the universal Regge slope, the
 tension in this case must be about $\frac{\pi}{4}$ times
the one from equation (\ref{eq:nambu}). Therefore, we fix $a$
to be $0.16\ GeV^{2}$, choose $m_{u,d}=0.3\ GeV$
and fit to the spin averaged
heavy-light meson states. Result is shown in Table \ref{tab:np}, and
parameters of the fit are
\bea
m_{u,d}&=&0.300 \ GeV \ {\rm (fixed)}\ ,\nonumber \\
m_{s}&=&0.600\ GeV\ ,\nonumber \\
m_{c}&=&1.342\ GeV\ ,\nonumber \\
m_{b}&=&4.679\ GeV\ ,\label{eq:npfit}\\
a&=&0.157 \ GeV^{2}\ {\rm (fixed)}\ ,\nonumber \\
\kappa&=&0.676\ . \nonumber
\eea
The agreement of the
fitted levels to experiment
is very good
and comparable to the Dirac scalar confinement fit of Table \ref{tab:sconf}.

Finally we use the wavefunctions
with parameters of  (\ref{eq:npfit}) to evaluate the IW function
using (\ref{eq:iwf}). The result is shown on  Fig. \ref{fig:npiwf}.
The slope
of the Isgur-Wise function at the zero recoil point
is evaluated using (\ref{eq:first}) to be
\be
\xi'(1)=-1.20\pm 0.03\ ,
\ee
where  the error is estimated from the variation of the light
quark mass $0.25<m_{u,d}<0.35\ GeV$. We observe that this slope is
significantly more negative than the one found from the
Dirac equation with scalar confinement (\ref{eq:scslope}), even though
the same set of spin averaged heavy-light meson masses were
used in the fit.
However, if we compare Figs. \ref{fig:sciwf} and \ref{fig:npiwf}, we
see that Isgur-Wise function obtained from the
NP equation agrees with the data a  bit better than the
one calculated from the Dirac equation with scalar confinement.

\section{Conclusions}
\label{sec:conc}

We have considered here the motion of a fermion
in a central field. The interaction that we have emphasized
is scalar or vector linear confinement. We are particularly interested
in this problem because of its application to the description
of heavy-light mesons. The wave equations considered
are the Dirac equation and the ``no pair'' equation. Although the properties
of the Dirac equation are well known we reconsider them in the
light of a confining interaction and also to serve as benchmark
for the related NP equation. Our main results are:
\begin{enumerate}
\item Dirac equation with scalar linear confinement

This is the most straightforward confinement model
for a fermion. We find that it is a well posed problem
with a unique solution and that by adjusting quark masses, Coulomb constant
$\kappa$,
and  tension $a$, an excellent fit to heavy-light masses
can be found. The tension found ($a\simeq 0.31\ GeV$) is larger
than normally
obtained from heavy onia fits and this may be viewed as
evidence against scalar
confinement. The origin of this discrepancy is the scalar confinement
Regge slope $\alpha'=\frac{1}{2a}$. The slope is fixed
by the p-wave
heavy-light
states and yields a tension about 50\% larger than the usual value.
If we force the tension to  be $0.2 \ GeV^2$ then although the
fit to the energy levels is still good, the Isgur-Wise function
becomes shallow with a slope $\xi'(1)\simeq -0.70$ and does not
fit
the experimental data well. We conclude that although scalar
confinement gives mathematically consistent
solutions, it does not seem to agree well with experiment.

\item Dirac equation with vector confinement

There are no normalizable bound state solutions in this case.

\item NP equation with scalar confinement

Scalar confinement
in the no pair equation does not yield normal quasi-linear Regge
trajectories and hence does not correspond to
universal Regge behaviour.

\item NP equation with vector confinement

This model of fermionic confinement also is viable. Vector NP
confinement, along with an attractive Coulombic short
range interaction, again gives good fits to the data. Vector
confinement yields a Regge slope of
$\alpha'=\frac{1}{4a}$, one half of the scalar value.
\end{enumerate}

There are thus two alternative confinement models yielding
linear Regge trajectories:
scalar Dirac and vector NP.
They both account for the data well. The differences though
are interesting. The parameters, most notably the
tension, are different when fitted to the data.

As we observed in Section \ref{sec:regge} the Regge slope for scalar
and vector confinement differ by a factor of two. Since the
tension appears both in s-wave dynamics and in rotational states experimental
data will ultimately
decide the correct
result. The  tension prefered by heavy onia actually lies between
the scalar Dirac and vector NP heavy-light values.
As we have pointed out \cite{bib:collin,bib:sik} the
heavy-light Regge slope $\alpha'=\frac{1}{\pi a}$ obtained
in the relativistic flux tube model is consistent
with the heavy onia
value.

Another piece of information which should soon shed
additional light on the proper
model of confinement
is the Isgur-Wise function. Both by experiment and from the lattice
simulation
of QCD, accurate values of the IW function (or its slope at zero recoil point)
will be available. The Isgur-Wise function
appears to depend fairly sensitively on the confinement
model. As we have seen, Dirac scalar confinement
yields $\xi'(1)=-0.9$, while NP vector confinement
gives $\xi'(1)=-1.2$. Both of these models provide excellent
fits to the same data set.

Finally, we should mention
that we have been led into these questions by our investigation of the
relativistic
flux tube model \cite{bib:collin,bib:sik,bib:lacourse}. For low
orbital angular momentum states the flux
tube model is similar to vector confinement. It appears
from the results presented here
that
a NP type equation will be appropriate
for the flux tube with one light fermion.

\appendix

\section{Basis states}
\label{app:bstates}

The pseudo-Coulombic radial basis states are those used in previous
calculations
\cite{bib:jacobs,bib:collin,bib:sik},
\be
R_{il}(r)=N_{il}\beta^{\frac{3}{2}}(2\beta r)^{l}e^{-\beta r}
L_{i}^{2l+2}(2\beta r)\ ,
\ee
where
\be
N_{il}=\sqrt{\frac{8 (i!)}{(i+2l+2)!}}\ .
\ee
In these equations we assume $0\leq i \leq N-1$. For computational
precision and efficiency, the matrix representation of all operators
have been calculated analytically \cite{bib:ball}. For
$p_{r}^{2}=
-\frac{1}{r}\frac{\partial^{2}}{\partial r^{2}}r$,
$r$, $\frac{1}{r}$ and $\frac{1}{r^{2}}$ we have
\bea
\lan p_{r}^{2}\ran_{ij}\hspace{-7pt}
&=&
\beta^{2}\left[-\delta_{i,j}+2\frac{N_{jl}}{N_{il}}
\frac{-l(2l+3)(j-i)+(l+1)(2i+2l+3)}{(2l+1)(2l+3)}\right]\ ,\\
\lan r\ran_{ij}
\hspace{-7pt}&=&\hspace{-7pt}
\frac{1}{2\beta}\left[ (2i+2l+3)\delta_{i,j}
-\sqrt{j(j+2l+2)}\delta_{i,j-1}
-\sqrt{i(i+2l+2)}\delta_{i,j+1}
\right]\hspace{-5pt}\ ,\\
\lan \frac{1}{r} \ran_{ij}&=&
\frac{\beta}{l+1}\frac{N_{jl}}{N_{il}}\ ,\\
\lan \frac{1}{r^{2}}\ran_{ij} &=&
\frac{2\beta^{2}}{(l+1)(2l+1)(2l+3)}\frac{N_{jl}}{N_{il}}
\left[(2l+1)(j-i) + 2j+2l+3\right]\ ,
\eea
 where we assume $i\leq j$, while results for
$i> j$ can be obtained by simple reflection due to the
symmetry of
operators. The only other matrix element used was
\be
\lan\frac{\partial}{\partial r}\ran_{ij}=
\frac{\beta}{l+1} \left\{
	\begin{array}{ccc}
	-(l+2)\frac{N_{jl}}{N_{il}} &,& i<j \\
	-1			    &,& i=j \\
	l \frac{N_{il}}{N_{jl}}	    &,& i>j
	\end{array}\right. \ .
\ee

\begin{center}
ACKNOWLEDGMENTS
\end{center}
This work was supported in part by the U.S. Department of Energy
under Contract Nos.  DE-AC02-76ER00881 and DE-AC05-84ER40150,
the National Science
Foundation under Grant No. HRD9154080,
and in part by the University
of Wisconsin Research Commitee with funds granted by the Wisconsin Alumni
Research Foundation.

\begin{table}
\normalsize
\begin{center}
TABLES
\end{center}
\caption{ Heavy-light spin averaged states.
Theoretical
results are obtained from the Dirac equation with
scalar confinement. Spin-averaged
masses are calculated in the usual way, by taking $\frac{3}{4}$
($\frac{5}{8}$) of
the triplet and $\frac{1}{4}$ ($\frac{3}{8}$) of the singlet mass for the
s(p)-waves). An estimate
for the $B_{s}^{*}$ was taken to be $5421\  MeV$, in order to make
splittings $B^{*}-B$ and $B_{s}^{*}-B_{s}$ the same. }
\vskip 0.2cm
\begin{tabular}{|lccccccc|}
\hline
\hline
         state
       & \multicolumn{2}{c}{spectroscopic label\hspace{+2mm}}
       & spin-averaged
       & \multicolumn{2}{c}{q. n.}
       & theory
       & error
\\
       & $J^{P}$
       & $^{2S+1}L_{J}$
       & mass (MeV)
       & $j$
       & $k$
       & (MeV)
       & (MeV)
\\
\hline
         \underline{$c\bar{u},\ c\bar{d}$ quarks}
       &
       &
       &
       &
       &
       &
       &
\\
         $\begin{array}{ll}
              		D     &  (1867) \\
   			D^{*} &  (2010) \end{array}$
       & $\begin{array}{l}
       			0^{-} \\
      			1^{-} \end{array}$
       &
	 $\left. \begin{array}{l}
			\hspace{+1.1mm}   ^{1}S_{0} \\
			\hspace{+1.1mm}   ^{3}S_{1} \end{array}\right] $
       & $1S\ (1974)$
       & $\frac{1}{2}$
       & $-1$
       & $1975$
       & $1$
\\
	 $\begin{array}{ll}
			D_{1}     & (2423) \\
			D_{2}^{*} & (2457) \end{array}$
       & $\begin{array}{l}
			1^{+} \\
			2^{+} \end{array}$
       & $\left.\begin{array}{l}
			\hspace{+0.5mm}^{1}P_{1} \\
			\hspace{+0.5mm}^{3}P_{2} \end{array}\right] $
       & $1P\ (2444)$
       & $\frac{3}{2}$
       & $-2$
       & $2444$
       & $0$
\\
         \underline{$c\bar{s}$ quarks}
       &
       &
       &
       &
       &
       &
       &
\\
	 $\begin{array}{ll}
   			D_{s} & (1969) \\
   			D^{*}_{s}&  (2110) \end{array}$
       & $\begin{array}{l}
      			0^{-} \\
      			1^{-} \end{array}$
       & $\left. \begin{array}{l}
			\hspace{+1.1mm}    ^{1}S_{0} \\
			\hspace{+1.1mm}    ^{3}S_{1} \end{array}\right] $
       & $1S\ (2075)$
       & $\frac{1}{2}$
       & $-1$
       & $2074$
       & $-1$
\\
	 $\begin{array}{ll}
			D_{s_{1}} & (2535) \\
		        D_{s_{J}} & (2573) \end{array}$
       & $\begin{array}{l}
			1^{+} \\
			2^{+} \end{array}$
       & $\left.\begin{array}{l}
			\hspace{+0.5mm}^{1}P_{1} \\
		  	\hspace{+0.5mm}^{3}P_{2} \end{array}\right] $
       & $1P\ (2559)$
       & $\frac{3}{2}$
       & $-2$
       & $2559$
       & $0$
\\
         \underline{$b\bar{u},\ b\bar{d}$ quarks}
       &
       &
       &
       &
       &
       &
       &
\\
         $\begin{array}{ll}
   			B     & (5279) \\
   			B^{*} &  (5325) \end{array}$
       & $\begin{array}{l}
     	 		0^{-} \\
      			1^{-} \end{array}$
       & $\left. \begin{array}{l}
			\hspace{+1.1mm}   ^{1}S_{0} \\
 			\hspace{+1.1mm}   ^{3}S_{1} \end{array}\right] $
       & $1S\ (5314)$
       & $\frac{1}{2}$
       & $-1$
       & $5313$
       & $-1$
\\
         \underline{$b\bar{s}$ quarks}
       &
       &
       &
       &
       &
       &
       &
\\
         $\begin{array}{ll}
			B_{s}     & (5375) \\
			?B_{s}^{*} & (5421) \end{array}$
       & $\begin{array}{l}
			0^{-} \\
			1^{-} \end{array} $
       & $\left.\begin{array}{l}
			\hspace{+1.0mm} ^{1}S_{0} \\
			\hspace{+1.0mm} ^{3}S_{1} \end{array}\right] $
       & $1S\ (5410)$
       & $\frac{1}{2}$
       & $-1$
       & $5411$
       & $1$
\\
\hline
\hline
\end{tabular}
\\
\label{tab:sconf}
\end{table}

\begin{table}
\caption{ Heavy-light spin averaged states.
Theoretical
results are obtained from the no pair equation with
vector confinement. Spin-averaged
masses are calculated in the usual way, by taking $\frac{3}{4}$
($\frac{5}{8}$) of
the triplet and $\frac{1}{4}$ ($\frac{3}{8}$) of the singlet mass for the
s(p)-waves). An estimate
for the $B_{s}^{*}$ was taken to be $5421\  MeV$, in order to make
splittings $B^{*}-B$ and $B_{s}^{*}-B_{s}$ the same.}
\vskip 0.2cm
\begin{tabular}{|lccccccc|}
\hline
\hline
         state
       & \multicolumn{2}{c}{spectroscopic label\hspace{+2mm}}
       & spin-averaged
       & \multicolumn{2}{c}{q. n.}
       & theory
       & error
\\
       & $J^{P}$
       & $^{2S+1}L_{J}$
       & mass (MeV)
       & $j$
       & $k$
       & (MeV)
       & (MeV)
\\
\hline
         \underline{$c\bar{u},\ c\bar{d}$ quarks}
       &
       &
       &
       &
       &
       &
       &
\\
         $\begin{array}{ll}
              		D     &  (1867) \\
   			D^{*} &  (2010) \end{array}$
       & $\begin{array}{l}
       			0^{-} \\
      			1^{-} \end{array}$
       &
	 $\left. \begin{array}{l}
			\hspace{+1.1mm}   ^{1}S_{0} \\
			\hspace{+1.1mm}   ^{3}S_{1} \end{array}\right] $
       & $1S\ (1974)$
       & $\frac{1}{2}$
       & $-1$
       & $1978$
       & $4$
\\
	 $\begin{array}{ll}
			D_{1}     & (2423) \\
			D_{2}^{*} & (2457) \end{array}$
       & $\begin{array}{l}
			1^{+} \\
			2^{+} \end{array}$
       & $\left.\begin{array}{l}
			\hspace{+0.5mm}^{1}P_{1} \\
			\hspace{+0.5mm}^{3}P_{2} \end{array}\right] $
       & $1P\ (2444)$
       & $\frac{3}{2}$
       & $-2$
       & $2440$
       & $-4$
\\
         \underline{$c\bar{s}$ quarks}
       &
       &
       &
       &
       &
       &
       &
\\
	 $\begin{array}{ll}
   			D_{s} & (1969) \\
   			D^{*}_{s}&  (2110) \end{array}$
       & $\begin{array}{l}
      			0^{-} \\
      			1^{-} \end{array}$
       & $\left. \begin{array}{l}
			\hspace{+1.1mm}    ^{1}S_{0} \\
			\hspace{+1.1mm}    ^{3}S_{1} \end{array}\right] $
       & $1S\ (2075)$
       & $\frac{1}{2}$
       & $-1$
       & $2072$
       & $-3$
\\
	 $\begin{array}{ll}
			D_{s_{1}} & (2535) \\
		        D_{s_{J}} & (2573) \end{array}$
       & $\begin{array}{l}
			1^{+} \\
			2^{+} \end{array}$
       & $\left.\begin{array}{l}
			\hspace{+0.5mm}^{1}P_{1} \\
		  	\hspace{+0.5mm}^{3}P_{2} \end{array}\right] $
       & $1P\ (2559)$
       & $\frac{3}{2}$
       & $-2$
       & $2562$
       & $3$
\\
         \underline{$b\bar{u},\ b\bar{d}$ quarks}
       &
       &
       &
       &
       &
       &
       &
\\
         $\begin{array}{ll}
   			B     & (5279) \\
   			B^{*} &  (5325) \end{array}$
       & $\begin{array}{l}
     	 		0^{-} \\
      			1^{-} \end{array}$
       & $\left. \begin{array}{l}
			\hspace{+1.1mm}   ^{1}S_{0} \\
 			\hspace{+1.1mm}   ^{3}S_{1} \end{array}\right] $
       & $1S\ (5314)$
       & $\frac{1}{2}$
       & $-1$
       & $5315$
       & $1$
\\
         \underline{$b\bar{s}$ quarks}
       &
       &
       &
       &
       &
       &
       &
\\
         $\begin{array}{ll}
			B_{s}     & (5375) \\
			?B_{s}^{*} & (5421) \end{array}$
       & $\begin{array}{l}
			0^{-} \\
			1^{-} \end{array} $
       & $\left.\begin{array}{l}
			\hspace{+1.0mm} ^{1}S_{0} \\
			\hspace{+1.0mm} ^{3}S_{1} \end{array}\right] $
       & $1S\ (5410)$
       & $\frac{1}{2}$
       & $-1$
       & $5409$
       & $-1$
\\
\hline
\hline
\end{tabular}
\\
\label{tab:np}
\vspace*{+2cm}
\end{table}

\begin{figure}[p]
\begin{center}
FIGURES
\vskip 2mm
\end{center}
\caption{S-wave classical turning points for the
Dirac equation with scalar confinement.}
\label{fig:tps}
\end{figure}

\begin{figure}
\caption{S-wave classical turning points for the
Dirac equation with vector confinement.}
\label{fig:tpv}
\end{figure}

\begin{figure}
\caption{S-wave classical turning points for the
no pair equation. Note that there is no difference
between the scalar and the vector confinement.}
\label{fig:tpnp}
\end{figure}

\begin{figure}
\caption{Dependence on $\beta$ of the three lowest positive energy states for
the pure Coulomb case with
$m=1\ GeV,\ \kappa=0.5,\ k=-1$ and $j=\frac{1}{2}$. Exact
solutions of the Dirac equation are shown by full lines, while
our variational solutions are shown by dashed lines
(shorter for $N=15$ and longer for $N=25$ basis states used).}
\label{fig:cbplt}
\end{figure}

\begin{figure}
\caption{Scaled eigenvalue $\varepsilon$ for the ground
state ($j=\frac{1}{2},\ k=-1$). We used $m=1\ GeV$ and
$N=25$ basis states. The no pair equation result is shown
by the dashed line, while the Dirac equation result is shown by
the full line.}
\label{fig:kplt}
\end{figure}

\begin{figure}
\caption{Dependence on $\beta$ of the three lowest positive energy states
and the three highest negative energy states for
the Dirac equation with scalar confinement, with
$m_{q}=0.3\ GeV,\ a=0.2\ GeV^{2},
\ \kappa=0.5,\ k=-1$ and $j=\frac{1}{2}$. Full lines correspond to
$N=25$, and dashed lines correspond to
$N=15$
basis states used.}
\label{fig:scbplt}
\end{figure}

\begin{figure}
\caption{Regge trajectories for the Dirac equation with scalar confinement.
We have chosen $m_{u,d}=0,\ a=0.2 GeV^{2}$, and
$\kappa = 0.5$. Full lines correspond to $k=-(j+\frac{1}{2})$,
and dashed lines to $k=j+\frac{1}{2}$. To ensure
that all calculated energies are correct, we used $N=100$ basis
states, and kept first 15 states.}
\label{fig:scregge}
\end{figure}

\begin{figure}
\caption{IW function for $\bar{B}$ decays calculated from the
Dirac equation with scalar confinement. Values for the
light quark mass $m_{u,d}$,  tension $a$ and
short range potential constant $\kappa$ are taken from the
 (\protect\ref{eq:scfit}). For the
sake of clarity, error bars are shown only for the
CLEO data.}
\label{fig:sciwf}
\end{figure}

\begin{figure}
\caption{Dependence on $\beta$ of the three lowest positive energy states
and the three highest negative energy states for
the Dirac equation with vector confinement, with
$m_{q}=0.3\ GeV,\ a=0.2\ GeV^{2},
\ \kappa=0.5,\ k=-1$ and $j=\frac{1}{2}$. We used
$N=15$
basis states. The mixing between positive and negative states is evident.}
\label{fig:vcbplt}
\end{figure}

\begin{figure}
\caption{Dependence on $\beta$ of the three lowest positive energy states
and the three highest negative energy states for
the no pair equation with vector confinement, with
$m_{q}=0.3\ GeV,\ a=0.2\ GeV^{2},
\ \kappa=0.5,\ k=-1$ and $j=\frac{1}{2}$. Full lines correspond to
$N=25$, and dashed lines correspond to
$N=15$
basis states used.}
\label{fig:npbplt}
\end{figure}

\begin{figure}
\caption{Regge
trajectories for the no pair equation with vector confinement.
We have chosen $m_{u,d}=0.3\ GeV,\ a=0.2 GeV^{2}$, and
$\kappa = 0.5$. Full lines correspond to $k=-(j+\frac{1}{2})$,
and dashed lines to $k=j+\frac{1}{2}$. To ensure
that all calculated energies are correct, we used $N=50$ basis
states, and kept first 10 states.}
\label{fig:npregge}
\end{figure}

\begin{figure}
\caption{IW function for $\bar{B}$ decays calculated from the
no pair equation with vector confinement. Values for the
light quark mass $m_{u,d}$,  tension $a$ and
short range potential constant $\kappa$ are taken from
 (\protect\ref{eq:npfit}). For the
sake of clarity, error bars are shown only for the
CLEO data.}
\label{fig:npiwf}
\end{figure}

\begin{figure}
\vskip 3cm
\end{figure}

\clearpage

\begin{figure}[p]
\epsfxsize = 5.4in \epsfbox{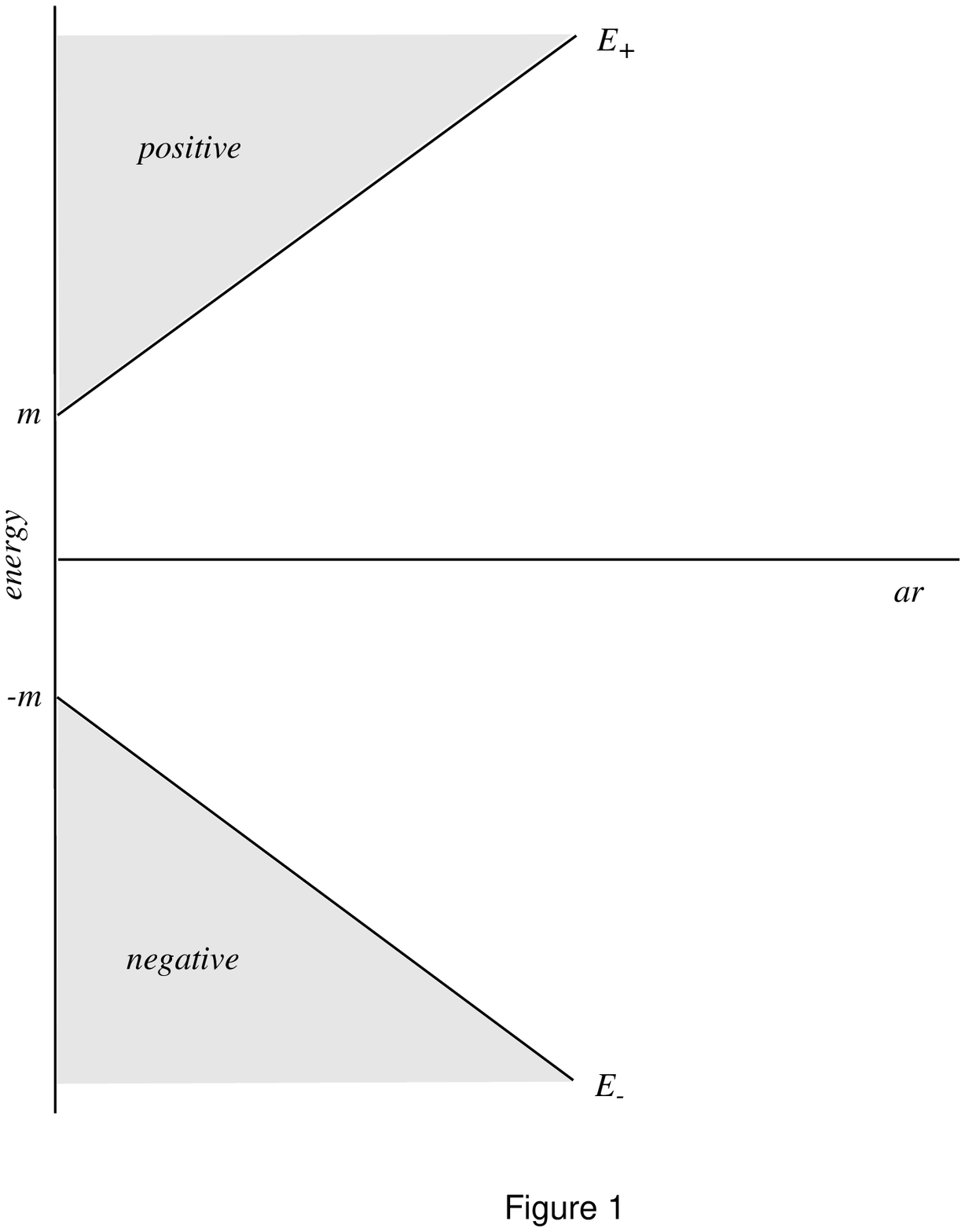}
\end{figure}

\begin{figure}[p]
\epsfxsize = 5.4in \epsfbox{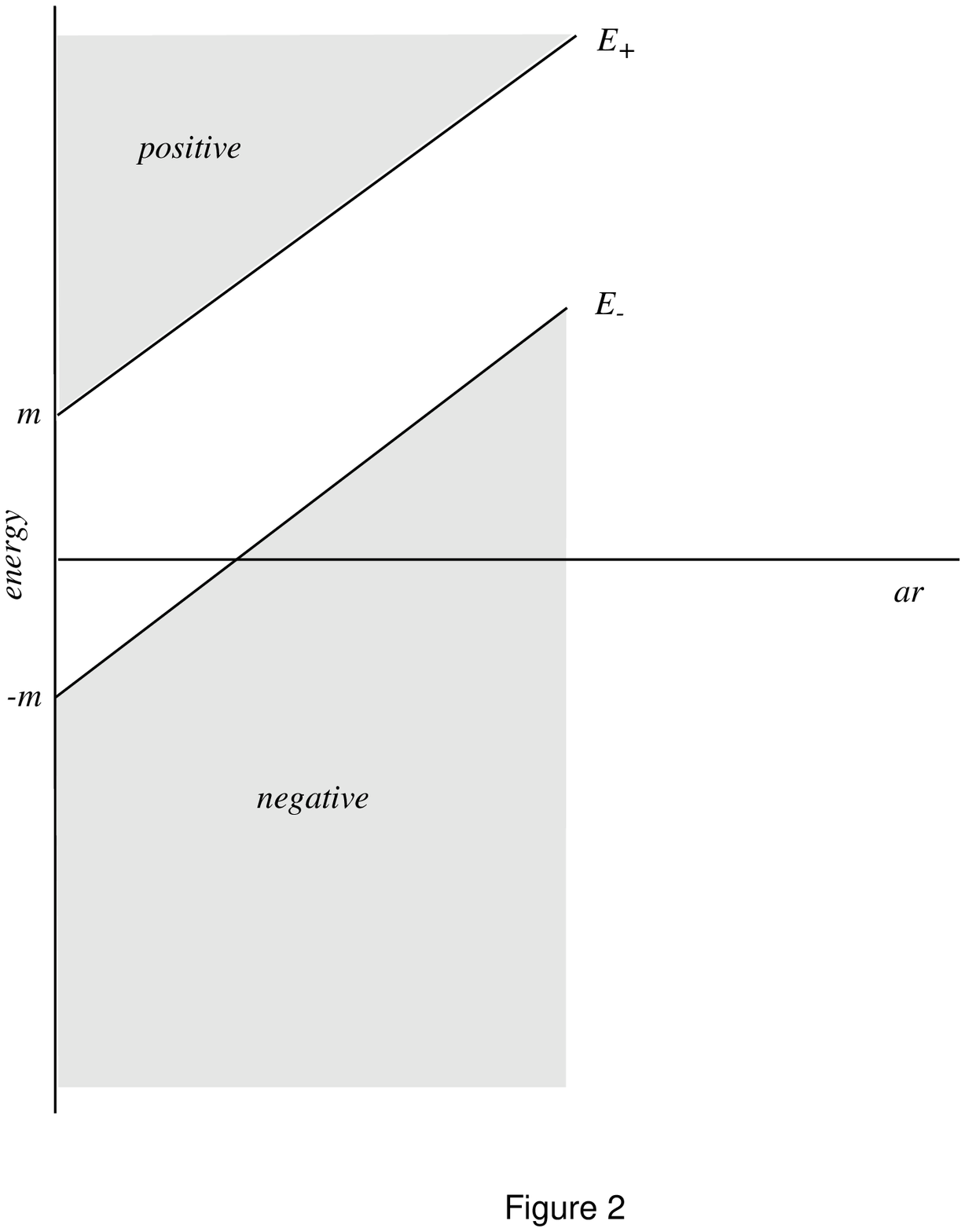}
\end{figure}

\begin{figure}[p]
\epsfxsize = 5.4in \epsfbox{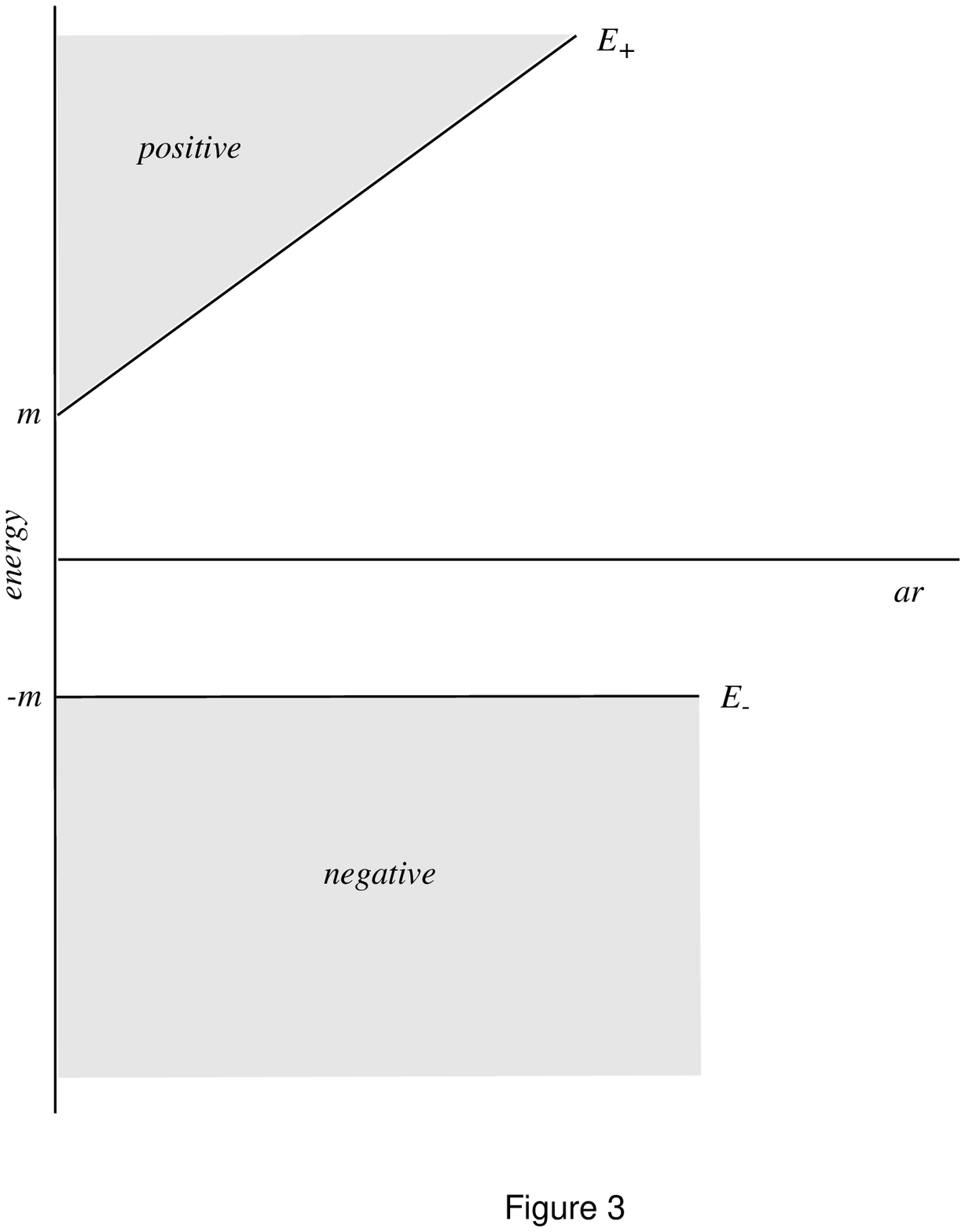}
\end{figure}

\begin{figure}[p]
\epsfxsize = 5.4in \epsfbox{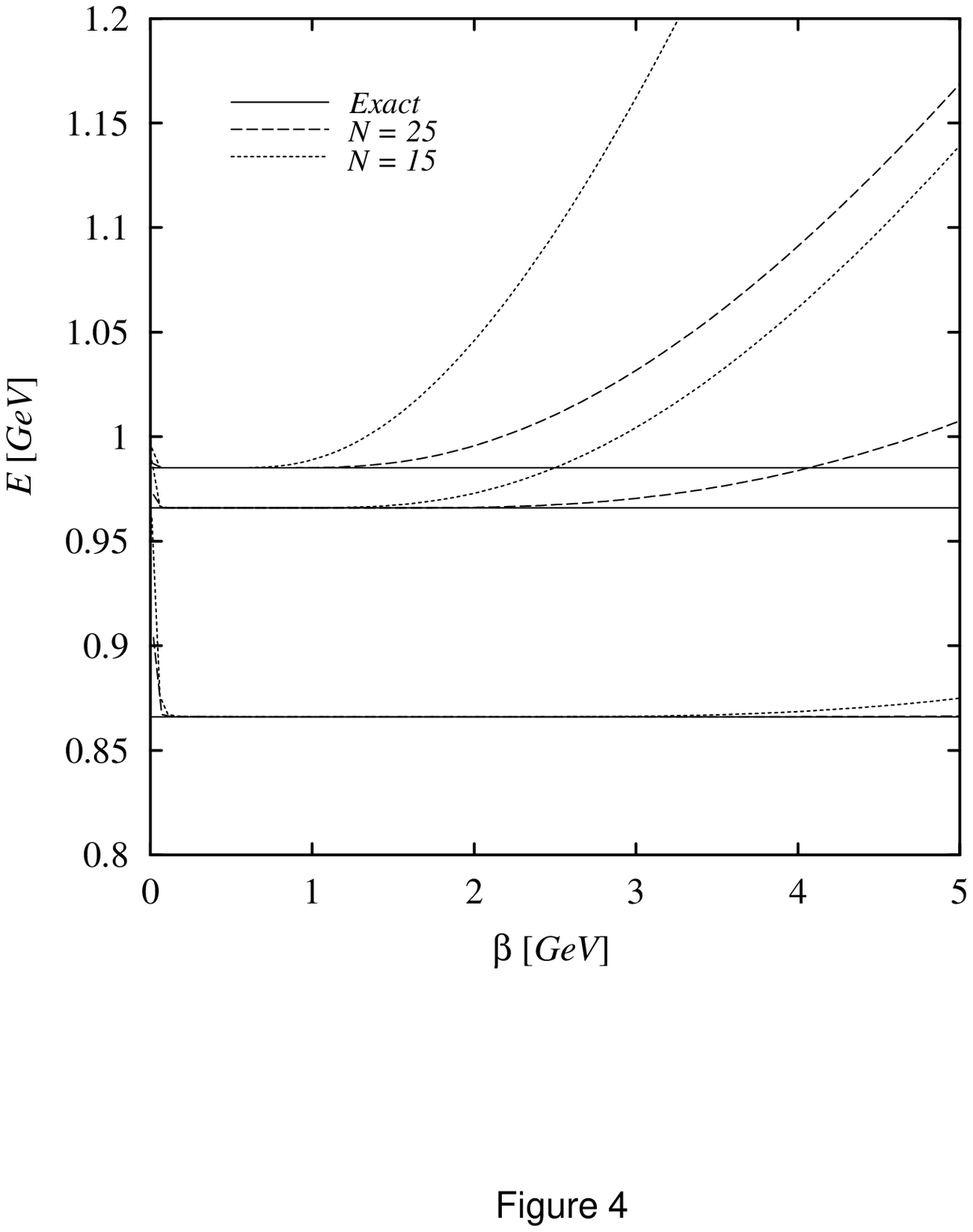}
\end{figure}

\begin{figure}[p]
\epsfxsize = 5.4in \epsfbox{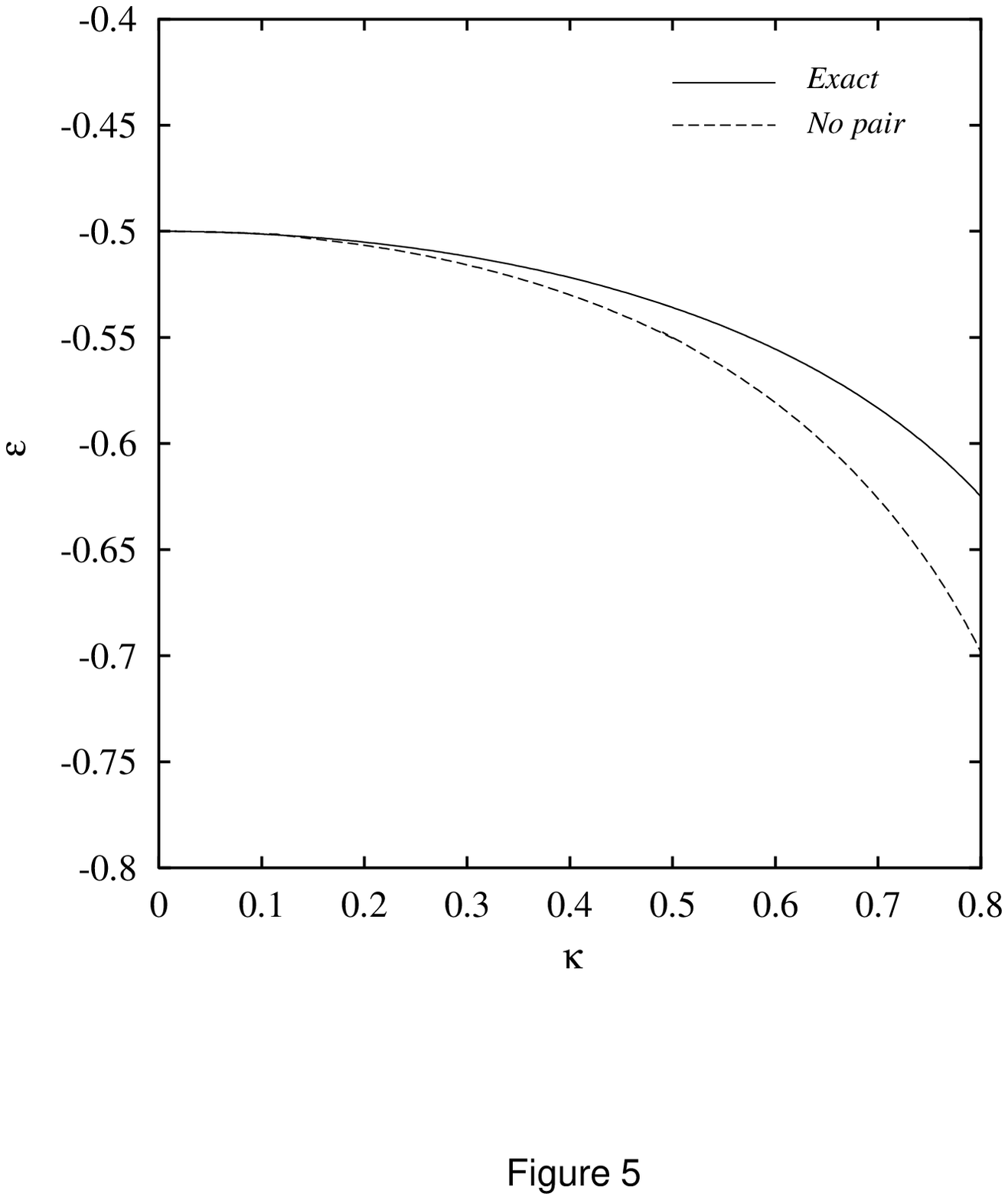}
\end{figure}

\begin{figure}[p]
\epsfxsize = 5.4in \epsfbox{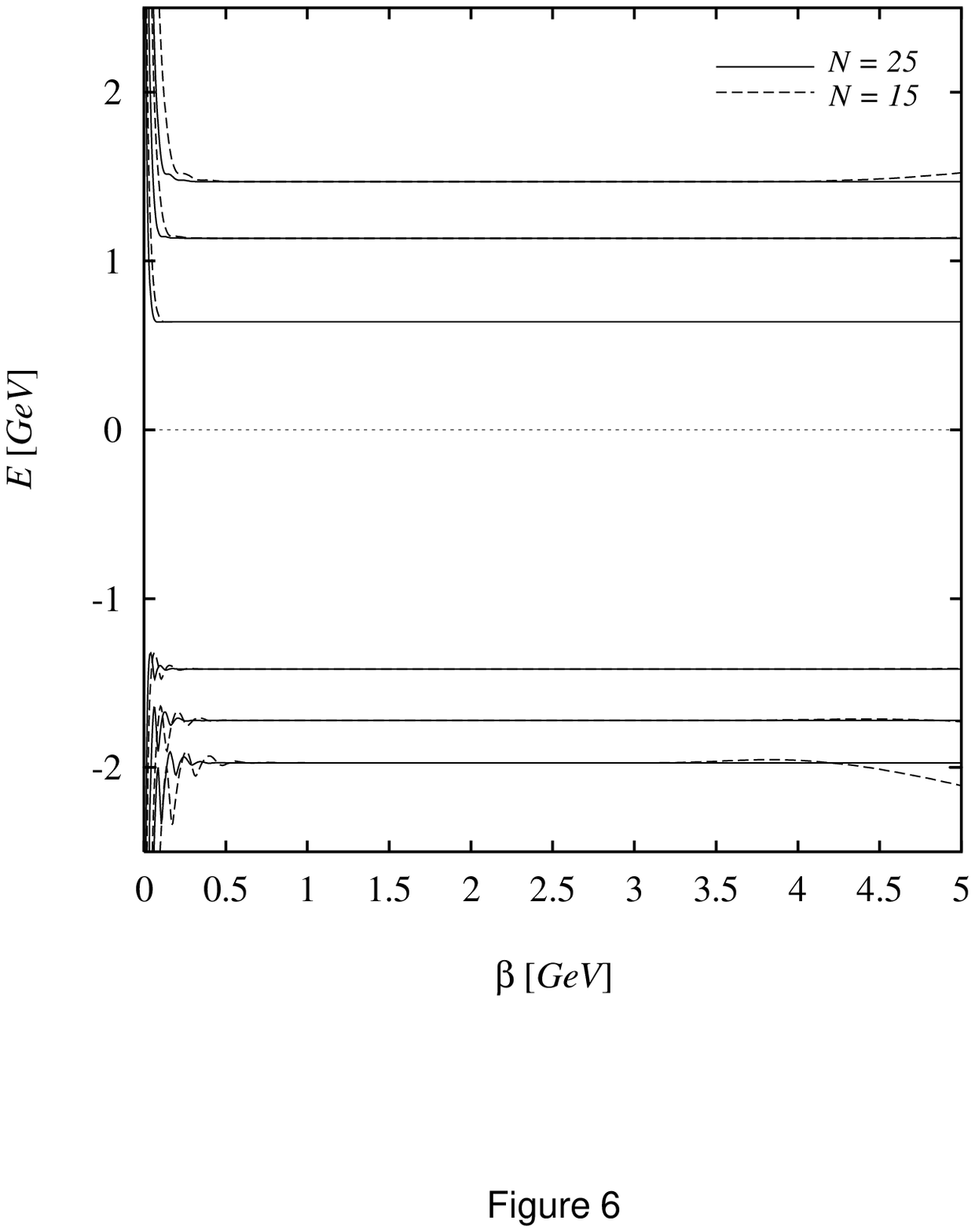}
\end{figure}

\begin{figure}[p]
\epsfxsize = 5.4in \epsfbox{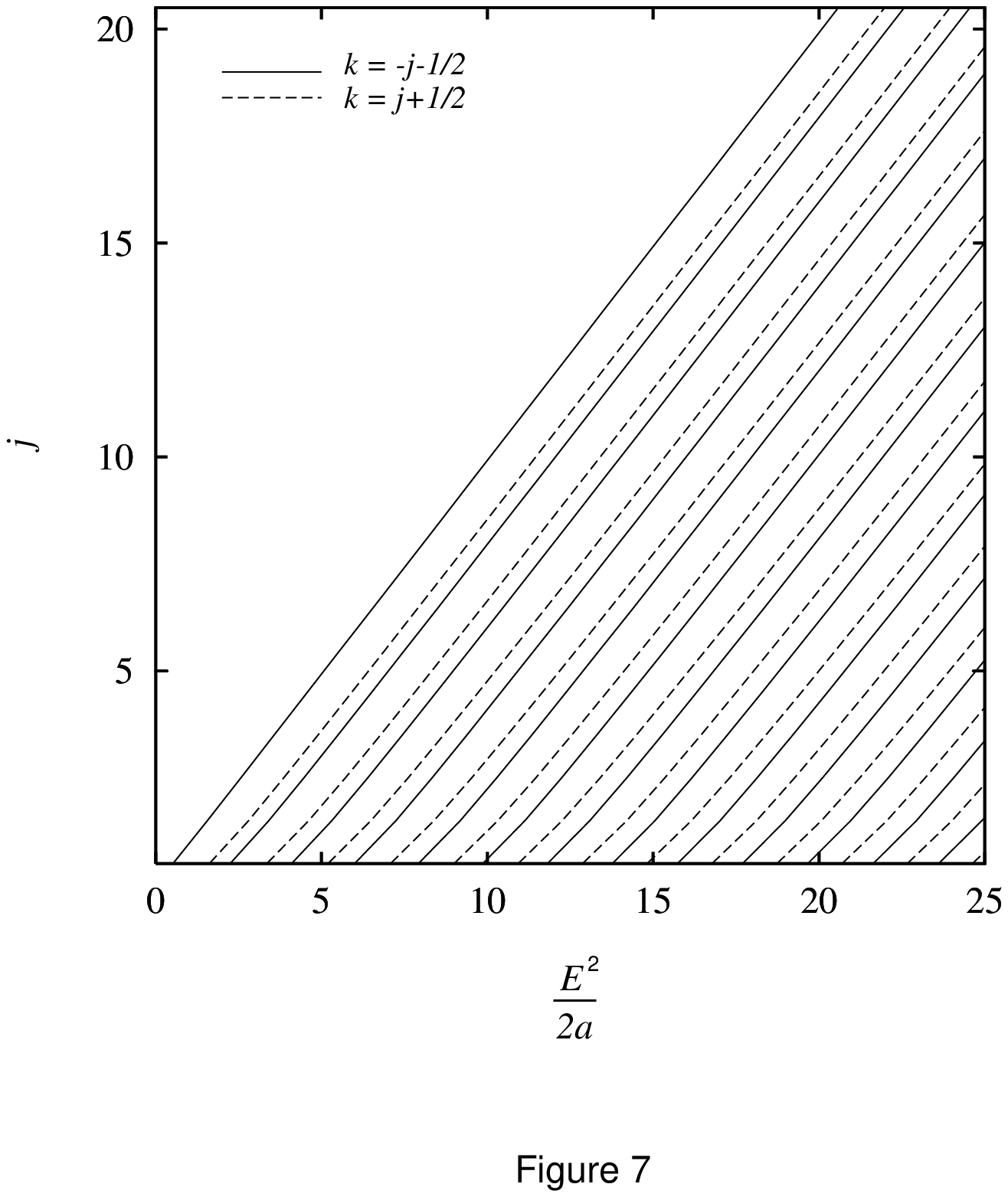}
\end{figure}

\begin{figure}[p]
\epsfxsize = 5.4in \epsfbox{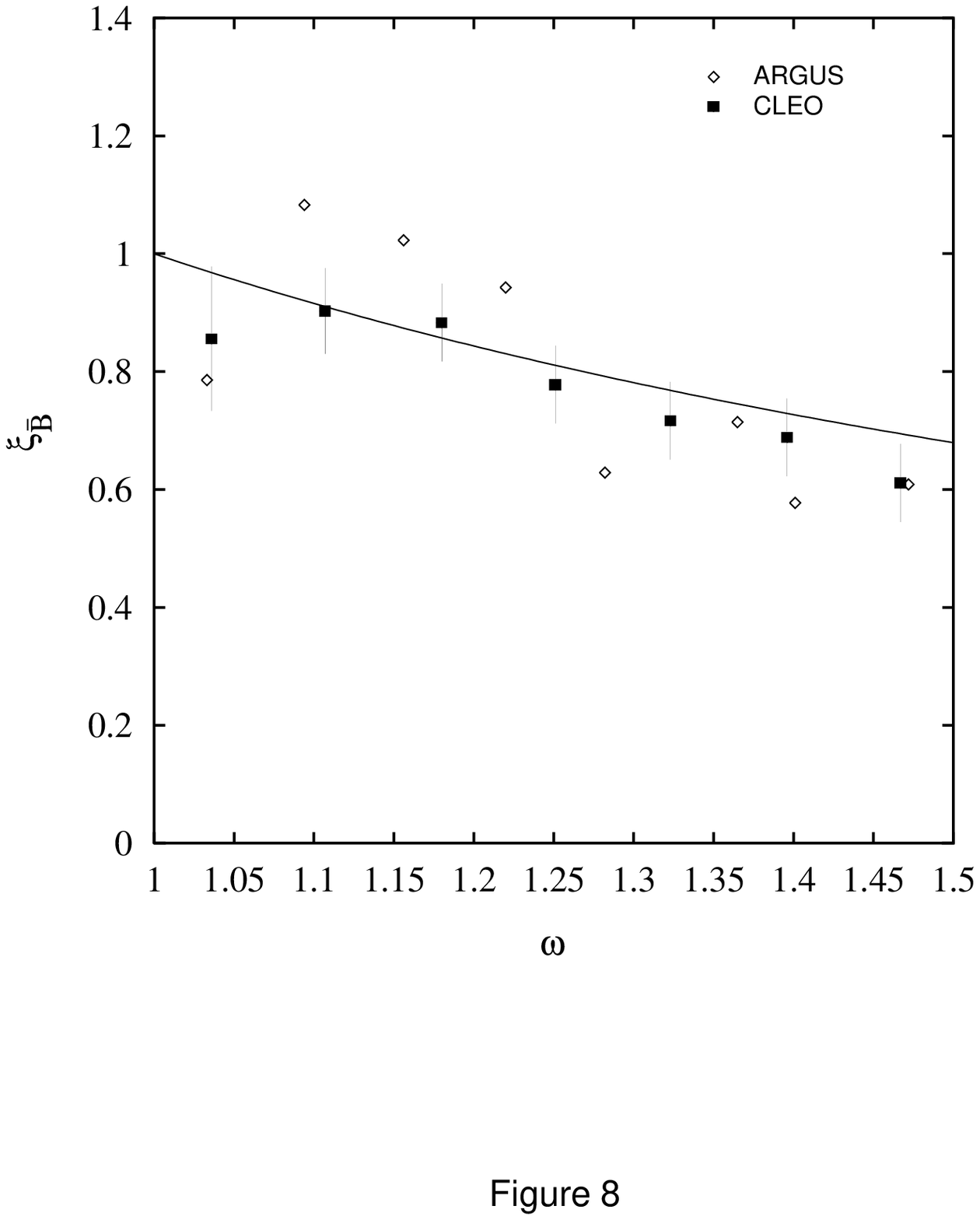}
\end{figure}

\begin{figure}[p]
\epsfxsize = 5.4in \epsfbox{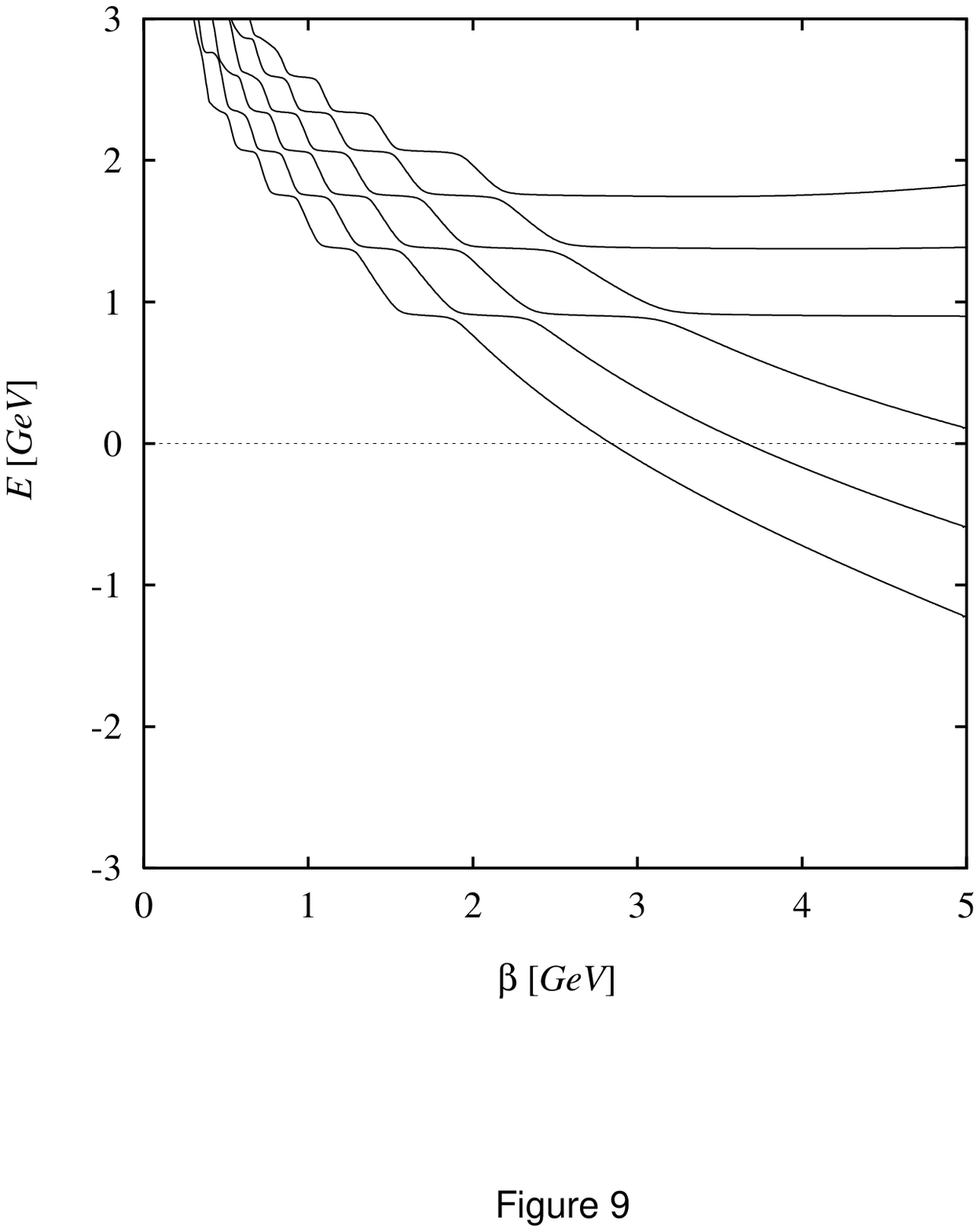}
\end{figure}

\begin{figure}[p]
\epsfxsize = 5.4in \epsfbox{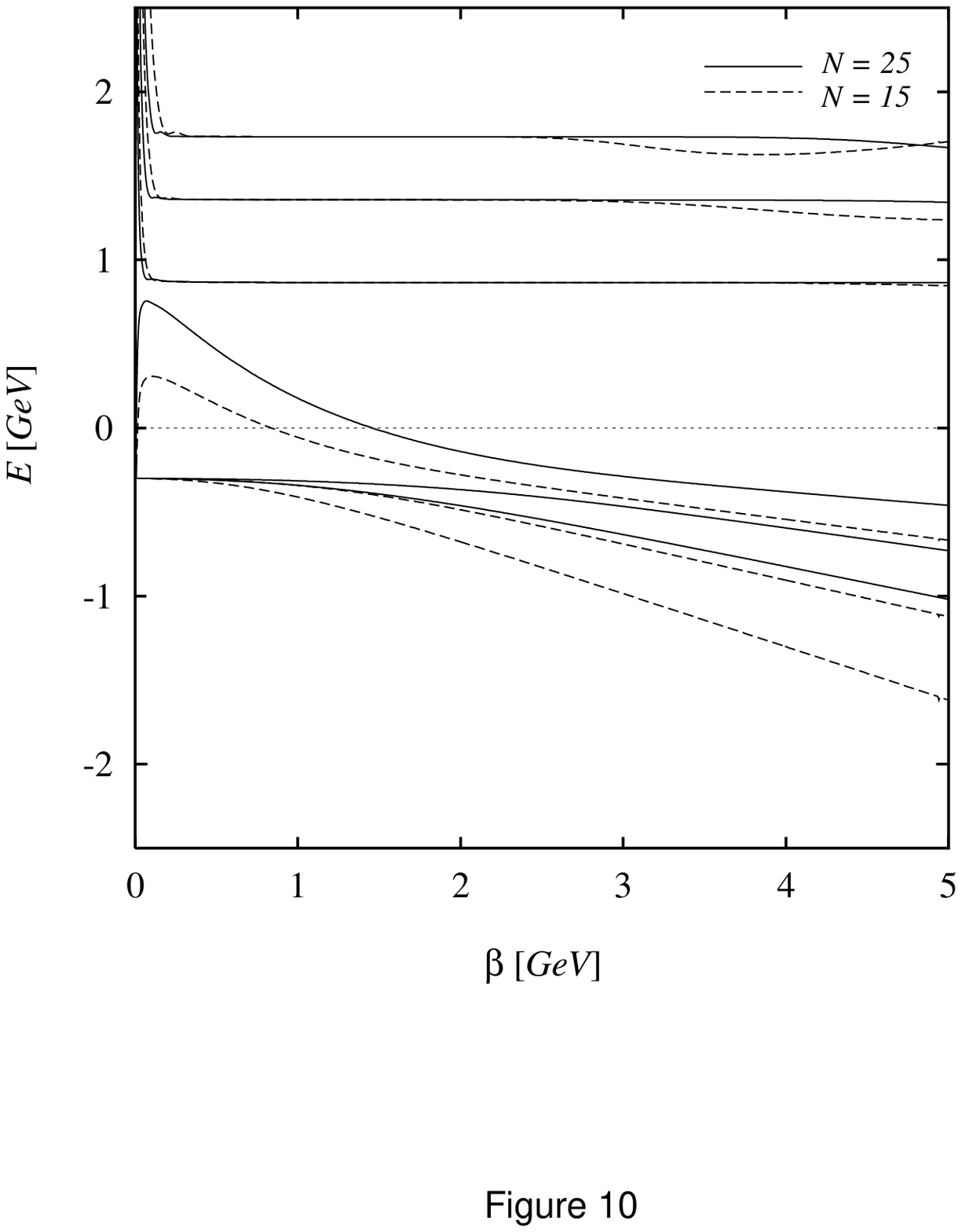}
\end{figure}

\begin{figure}[p]
\epsfxsize = 5.4in \epsfbox{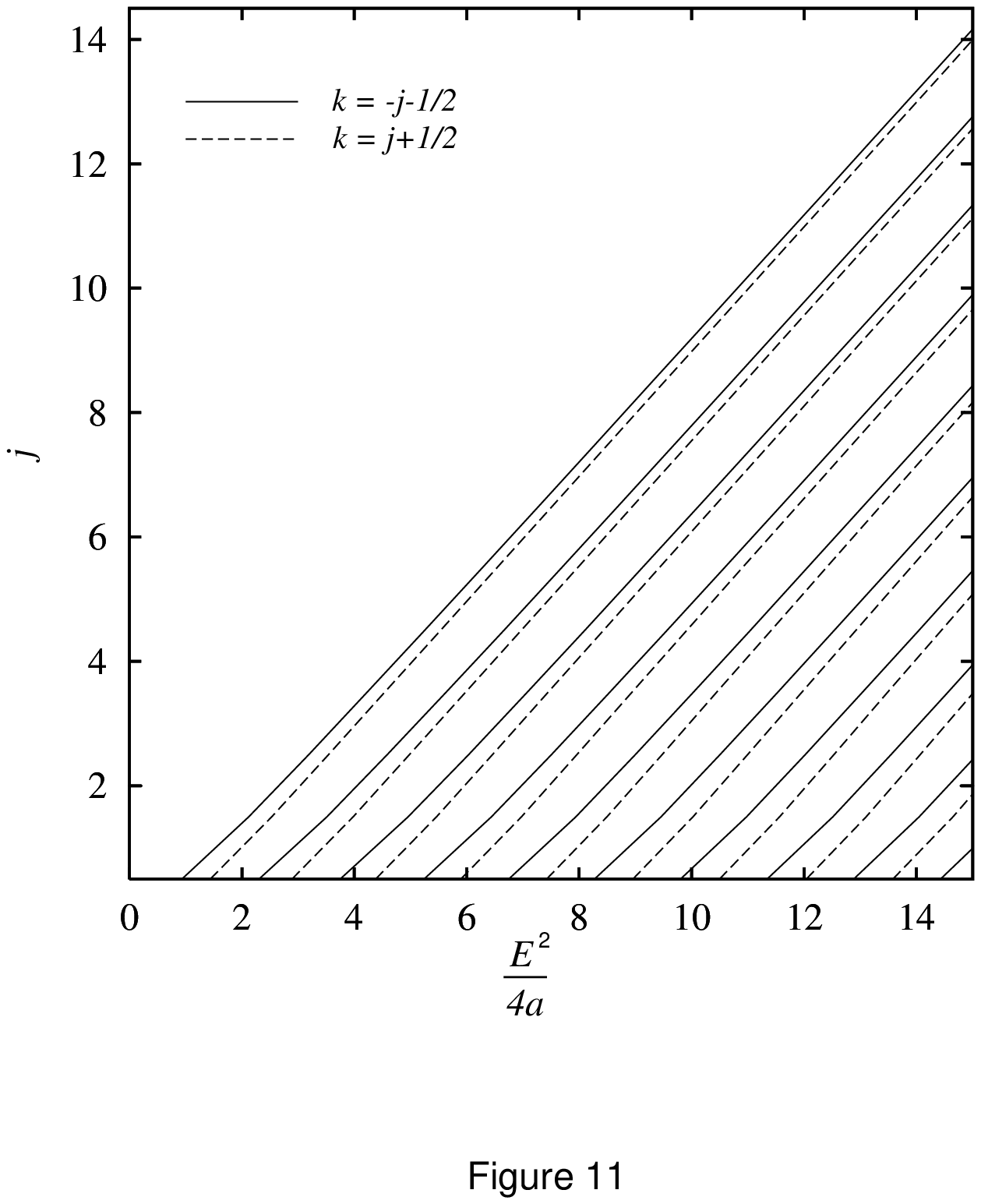}
\end{figure}

\begin{figure}[p]
\epsfxsize = 5.4in \epsfbox{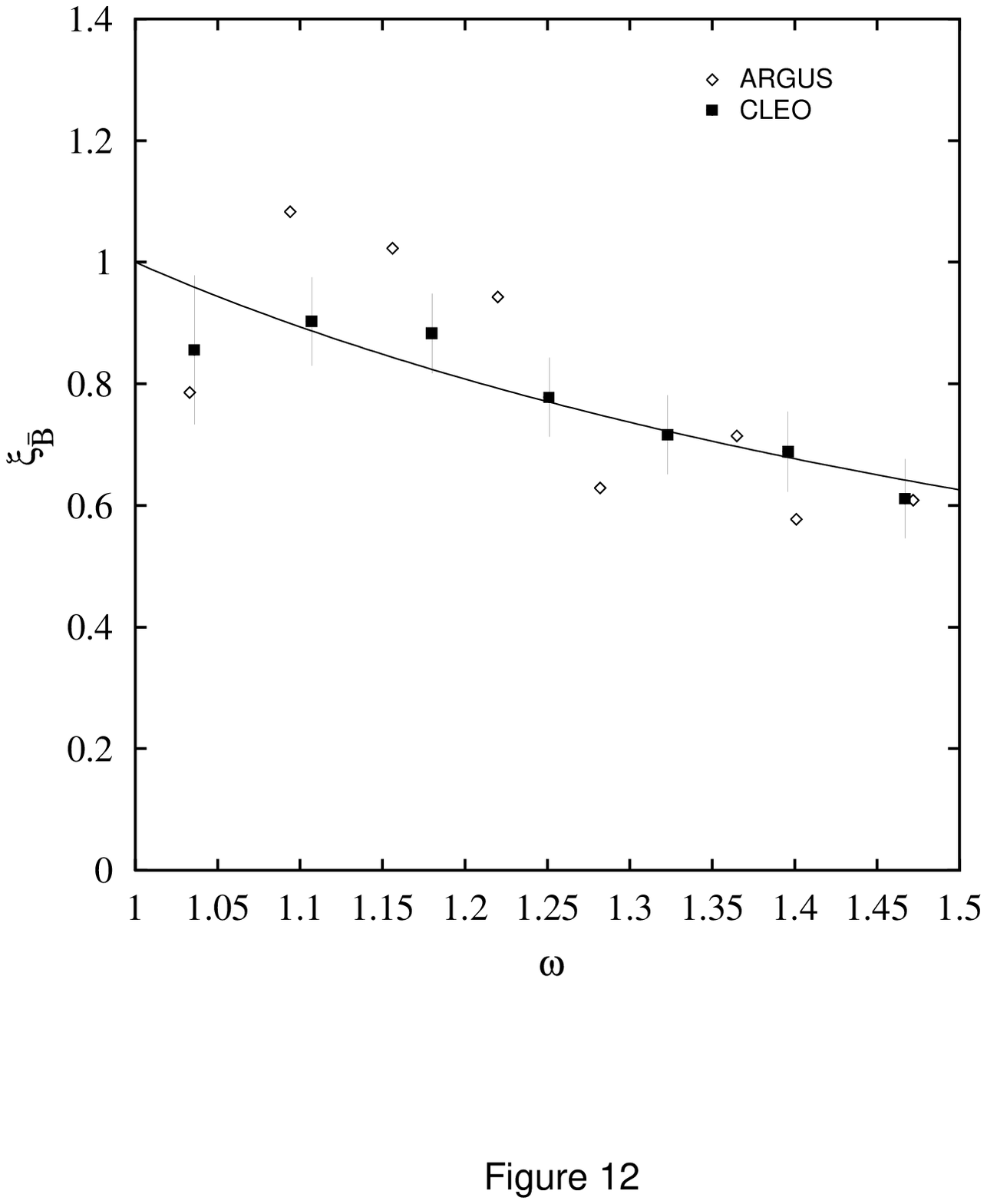}
\end{figure}

\end{document}